\documentclass[aps,prb,preprint,groupedaddress]{revtex4-1}
\usepackage{graphicx}
\usepackage{float}
\usepackage{amsmath}

\usepackage{amsbsy}
\usepackage[normalem]{ulem}

\newcommand{\bra}[1]{\langle #1|}
\newcommand{\ket}[1]{|#1 \rangle }
\newcommand{\beq}{\begin{equation}}
\newcommand{\eeq}{\end{equation}}
\newcommand{\beqn}{\begin{eqnarray}}
\newcommand{\eeqn}{\end{eqnarray}}

\begin{document}

\markboth{Pickard et al.}{Superconducting hydrides}

\title{Superconducting hydrides under pressure}

\author{Chris J.\ Pickard}
\email[]{cjp20@cam.ac.uk}
\affiliation{Department of Materials Science \& Metallurgy, University of Cambridge, 27 Charles Babbage Road, Cambridge CB3~0FS, United Kingdom}
\affiliation{Advanced Institute for Materials Research, Tohoku University 2-1-1 Katahira, Aoba, Sendai, 980-8577, Japan}

\author{Ion Errea}
\affiliation{Fisika Aplikatua 1 saila, Gipuzkoako Ingeniaritza Eskola, University of the Basque Country (UPV/EHU), Europa plaza 1, 20018 Donostia/San Sebasti\'an, Spain}
\affiliation{Centro de F\'isica de Materiales (CSIC-UPV/EHU), Manuel de Lardizabal 5, 20018 Donostia/San Sebasti\'an, Spain}
\affiliation{Donostia International Physics Center (DIPC), Manuel de Lardizabal 4, 20018 Donostia/San Sebasti\'an, Spain}

\author{Mikhail I. Eremets}
\affiliation{Max-Planck Institut f\"ur Chemie, Chemistry and Physics at High Pressures Group Postfach 3060, 55020 Mainz, Germany}

\begin{abstract}
The measurement of superconductivity at above 200K in compressed samples of hydrogen sulfide and lanthanum hydride at 250K is reinvigorating the search for conventional high temperature superconductors. At the same time it exposes a fascinating interplay between theory, computation and experiment. Conventional superconductivity is well understood, and theoretical tools are available for accurate predictions of the superconducting critical temperature. These predictions depend on knowing the microscopic structure of the material under consideration, and can now be provided through computational first principles structure predictions. The experiments at the megabar pressures required are extremely challenging, but for some groups at least, permit the experimental exploration of materials space. We discuss the prospects for the search for new superconductors, ideally at lower pressures.
\end{abstract}

\maketitle

\section{INTRODUCTION}

Kamerlingh Onnes's discovery in 1911 that mercury (Hg) abruptly begins to carry a current with no resistance at all when cooled below 4.2K~\cite{van2010discovery} was to puzzle for decades. Initially referred to as \emph{supraconductivity}, the temperature at which the resistance suddenly drops is now known as the \emph{superconducting} critical temperature, $T_c$. The new superconductors were found to completely exclude external magnetic fields by Meissner and Ochsenfeld in 1933~\cite{meissner1933neuer}. This is the \emph{Meissner effect} and, with no classical explanation, it is an essential hallmark of superconductivity. As well as high temperatures, high magnetic fields destroy the superconducting state. This critical field, $H_c$, is an important consideration for the technological application of superconductors. 

\begin{figure}[htbp]
\includegraphics[width=0.75\textwidth]{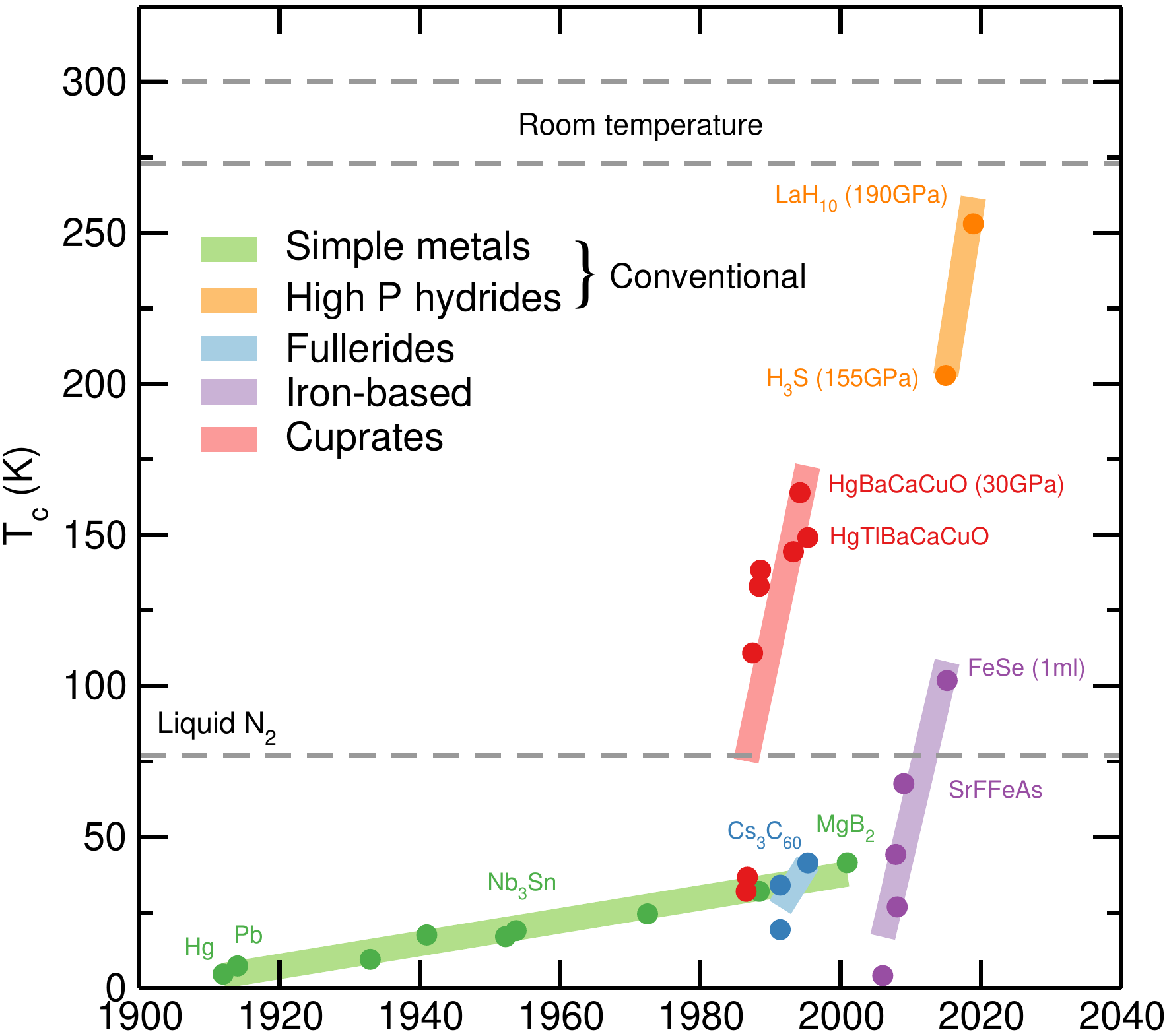}
\caption{Temporal evolution of the superconducting critical temperature $T_c$. Five families of superconductors are highlighted: the simple metals, fullerides, cuprates, iron-based, and high-pressure hydrides. The high-pressure hydrides are conventional superconductors as are the simple metals, while the cuprates and iron-based superconductors are unconventional. The pressure at which the measurement has been performed is given in parenthesis (if no value is provided it corresponds to ambient pressure). 1 ml stands for the monolayer case. Room and liquid nitrogen temperatures are indicated for reference.}
\label{fig1}
\end{figure}

Applications of superconductors include the generation of the intense magnetic fields required for magnetic resonance imaging (MRI) and particle accelerators, as well as superconducting quantum interference devices (SQUIDs), which are capable of measuring minute variations in magnetic fields. The applications are limited, however, by the extremely low temperatures that are needed to access the superconducting state (see Figure \ref{fig1}). The quest for high, or even room temperature superconductors has attained an iconic scientific status. In this review we describe the discovery of a new family of exceedingly high temperature superconductors - the high pressure hydrides.

\section{A DEVELOPING UNDERSTANDING}

From its discovery, superconductivity challenged the existing understanding of the behaviour of matter. It had not been (and could not have been using the theoretical tools then available) predicted beforehand. Soon lead (Pb) was found to superconduct at 7.2K~\cite{van2010discovery} and over the decades that followed many further superconducting materials were identified, culminating in the discovery in 1954 that Nb$_3$Sn superconducts with a $T_c$ of 18K~\cite{matthias1954superconductivity}. Importantly for the many applications that were to follow, Nb$_3$Sn could tolerate much higher external magnetic fields. 

The development of quantum mechanics in the 1920s supplied the missing theoretical tools, and a phenomenological theory of superconductivity emerged, most notably through the work of the London brothers~\cite{london1935electromagnetic}. But it would take some time, until the 1950s, before a microscopic picture of superconductivity could be pieced together. In 1950 it was discovered that superconductivity depended on the precise masses of the atoms involved~\cite{maxwell1950superconductivity,reynolds1950superconductivity}. This \emph{isotope effect} suggested to theorists that lattice vibrations, or phonons, play a central role in superconductivity. In 1957 Bardeen, Cooper and Schrieffer presented their microscopic theory of superconductivity~\cite{bardeen1957microscopic,bardeen1957theory}. In what would become known as BCS theory, the superconducting state is described in terms of \emph{Cooper pairs} of electrons, bound through the interaction between the electrons and phonons, which as bosons condense into a macroscopic quantum state. This theory  provides the basis of our understanding of what is now known as \emph{conventional} superconductivity. 

\section{SUPERCONDUCTING METALLIC HYDROGEN}

\begin{figure}[htbp]
\includegraphics[width=0.75\textwidth]{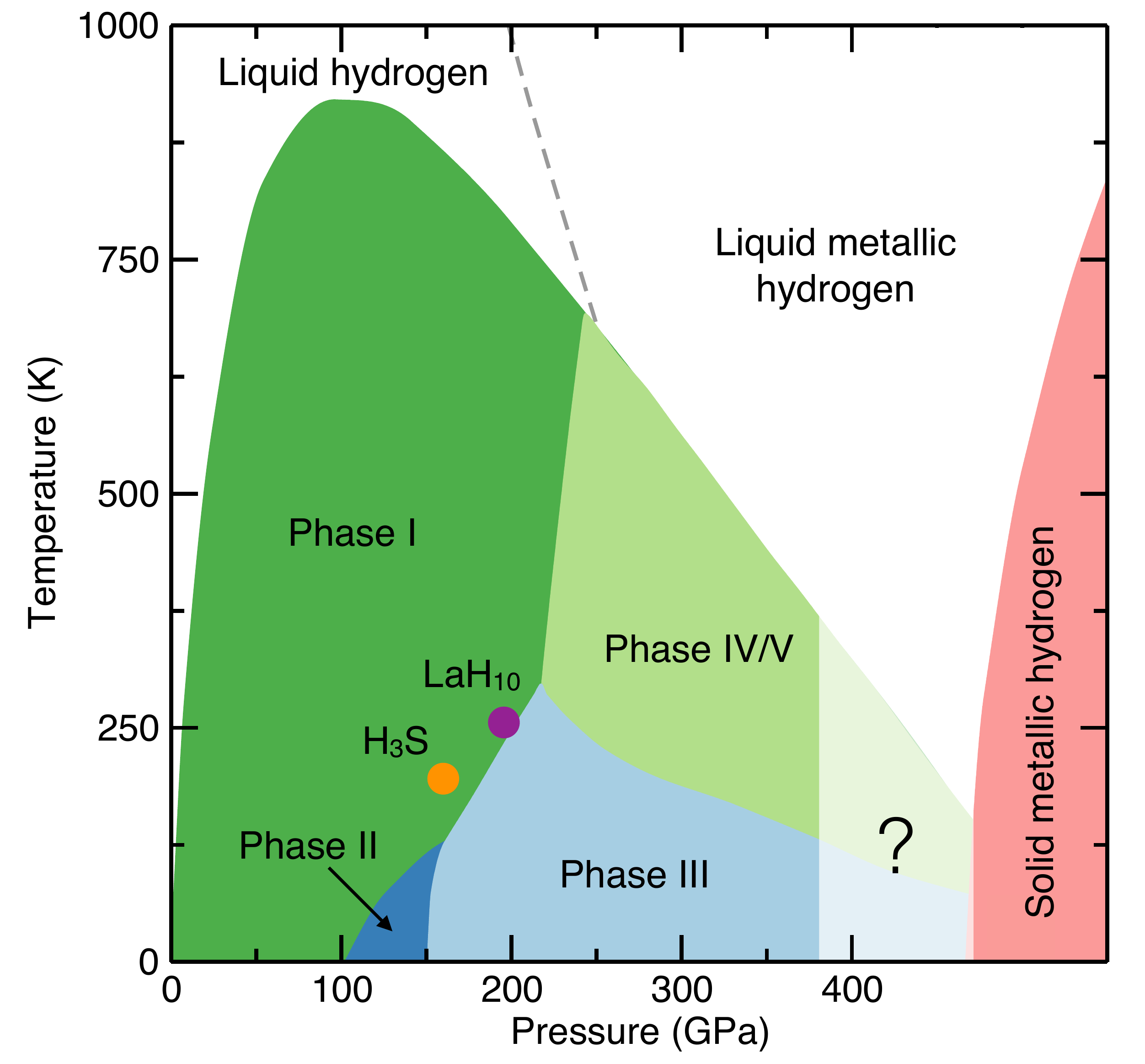}
\caption{Phase diagram of hydrogen. The superconducting transition temperatures and pressures for the experimentally observed superconducting hydrides, H$_3$S and LaH$_{10}$ are marked for reference. Microscopic models for the molecular phases have been provided by first principles structures predictions~\cite{pickard2007structure,mcmahon2012properties}, but the transition to solid metallic hydrogen is under intense experimental and theoretical scrutiny.}
\label{fig2}
\end{figure}

It has long been suspected that under sufficient compression hydrogen will join the Group I elements as a metal~\cite{wigner1935possibility}, and Figure \ref{fig2} summarises our current understanding of the phase diagram of hydrogen~\cite{mcmahon2012properties}. At low pressures, in phases I, II, and III, molecular hydrogen dominates. At high temperatures and pressures experiments find a metallic liquid phase (relevant to the gas giant planets in our solar system and beyond)~\cite{Zaghoo11873,Celliers677,Knudson1455}. However, at low temperatures there remains considerable controversy, even if recent optical measurements suggest a transition to solid metallic hydrogen at around 495 GPa~\cite{Dias715}. All theoretical results point to the existence of a solid metallic hydrogen phase at sufficient pressure~\cite{mcmahon2012properties}. This might be reached via a semi-metallic molecular phase~\cite{cudazzo:257001,monserrat2018structure}, or directly to an atomic phase~\cite{PhysRevLett.112.165501}. Both experiments and theoretical computations are extremely challenging in this transition regime.

Following the introduction of the BCS theory of conventional superconductivity, in 1968 Ashcroft proposed that solid metallic hydrogen, if it could be made, would be a high temperature superconductor~\cite{ashcroft1968metallic}. The BCS expression for the superconducting $T_c$ is:
\begin{equation}
    T_c=0.85\Theta_De^{-1/N(0)V}
\end{equation}
where $\Theta_D$ is the Debye temperature (derived from the highest frequency vibrational mode in the system), $N(0)$ is the density of electronic states at the Fermi level, and $V$ is an effective electron-phonon attractive interaction. The low mass of the proton ensures that metallic hydrogen will have a high Debye temperature, and assuming a reasonable value for $N(0)V$, Ashcroft predicted the $T_c$ to be very high.

While at that time metallic hydrogen was not within reach, Ashcroft's ideas were immediately put to work in the hunt for superconducting metallic hydrides at ambient pressures. However, the compounds investigated were either not superconducting, like the lanthanum hydrides~\cite{MERRIAM19631375,gupta1980electronic}, or superconducting with $T_c$ around 10 K~\cite{PhysRevLett.25.741,doi:10.1002/pssb.2220590133,Stritzker1972}. Little further progress was made, and attention was soon to be directed to a new class of superconductors.

\section{HIGH TEMPERATURE SUPERCONDUCTIVITY}

In 1986 research into superconductivity underwent a revolution due to the discovery of very high transition temperatures in a new class of materials -- the cuprates. Over a relatively short period of time the transition temperatures rocketed from around 30K in Ba$_x$La$_{5-x}$Cu$_5$O$_{5(3-y)}$, the result announced by Bednorz and M\"uller~\cite{bednorz1986possible}, to 164K in HgBaCaCuO~\cite{PhysRevB.50.4260} at 30GPa (see Figure \ref{fig1}). There was great optimism that room temperature superconductors were within our grasp. However, it soon became clear that these high temperature superconductors did not follow the same rules as the conventional BCS superconductors. These \emph{unconventional} superconductors demanded a new theoretical framework, one that despite intense effort and the deployment of many creative ideas~\cite{zaanen2010modern}, we still do not have. The cuprates have more recently been joined by the iron based superconductors~\cite{kamihara2006iron,kamihara2008iron}, but in the face of the diminishing increases in $T_c$ through doping or pressure, and despite providing a guide to the rich landscape of emergent phases in quantum matter~\cite{vojta2003quantum}, theory has not been in the position to provide a road-map to room temperature superconductivity based on these unconventional superconductors.

\section{NEW HOPE FOR THE CONVENTIONAL SUPERCONDUCTORS}

The discovery of the surprisingly high $T_c$, 39K, of MgB$_2$ in 2001~\cite{Nagamatsu2001} reminded the community of the potential of the conventional superconductors. The low cost of MgB$_2$ has made it an important superconductor for applications~\cite{doi:10.1111/j.1744-7402.2007.02138.x}. However, it appears to have been an isolated success, and subsequently discovered conventional superconductors (such as CaC$_6$~\cite{Weller2005}) have not surpassed it.

\begin{figure}[htbp]
\includegraphics[width=0.75\textwidth]{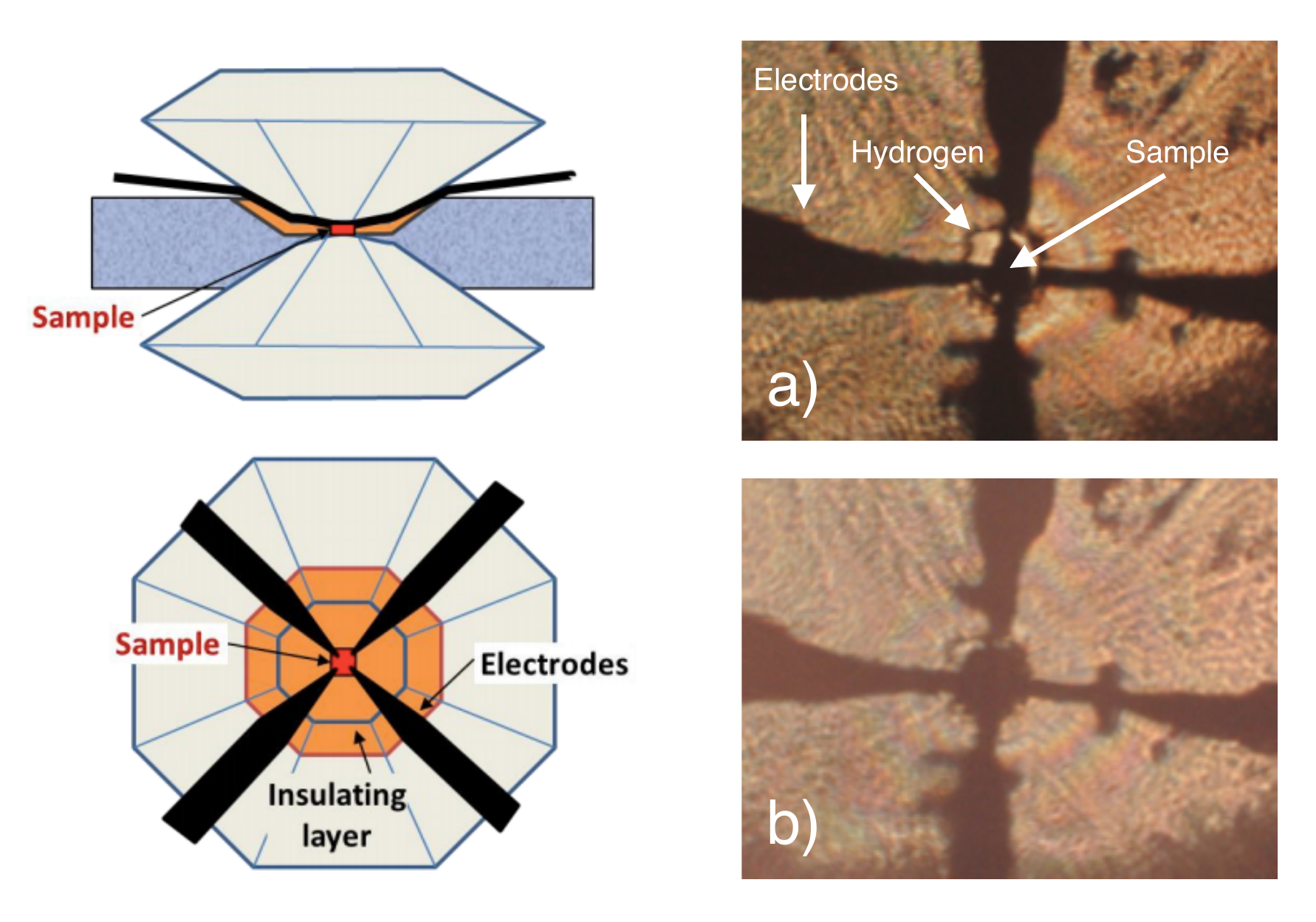}
\caption{Experimental synthesis and characterisation of dense hydrides in (Left) a diamond anvil cell (DAC). Adapted from a figure in Ref~\cite{Capitani2017}. (Right) Photographs of the Lanthanum hydride sample at 143 GPa (pressure determined from the shift of the Raman active vibron peak of hydrogen surrounding the sample). (a) Before laser heating, and (b) after laser heating the sample strongly expanded due to absorption of hydrogen.}
\label{fig3}
\end{figure}

In 2004 Ashcroft returned to his earlier ideas, this time explicitly suggesting that compounds with a high hydrogen content might be considered to be, in effect, chemically pre-compressed metallic hydrogen~\cite{ashcroft2004hydrogen}. With Hoffmann in 2006, a concrete proposal was made~\cite{PhysRevLett.96.017006}, and the era of a theory and computation led hunt for high temperature superconductors was upon us. The three developments that were central to this were 1) the reliable prediction of the stable structures of the hydrides under pressure, 2) the accurate computation of their superconducting properties, and 3) their experimental realization in diamond anvill cells (DACs) (see Figure \ref{fig3}). We will focus on the interplay between experiment, theory and computation that have together led to the new class of superconducting materials - the dense hydrides.

\section{DENSITY FUNCTIONAL THEORY}

In principle, solving the equations governing quantum physics -- Schrodinger or Dirac's equations -- would allow us to anticipate  the nature and properties of any material under conditions of our choice. In practice this is too difficult, or computationally expensive. It is typically assumed that the atomic nuclei are so much more massive than the electrons that the Born-Oppenheimer approximation holds, meaning only the lighter electrons need be treated quantum mechanically. This simplifies computation, considerably, but for the hydrides (because of the low mass of hydrogen) this approximation can break down. 

With Hohenberg, Kohn showed that the electronic charge density was sufficient to determine the ground state energy of a system~\cite{PhysRev.136.B864}. This energy can be written as a functional (or function of a function) of the density, hence Density Functional Theory (DFT). This put the earlier ideas of Thomas and Fermi on a solid theoretical footing. But the exact form of the appropriate functional remained, and remains, unknown. To create a useful computational scheme Kohn and Sham rewrote the charge density in terms of a set of functions reminiscent of independent particle orbitals~\cite{PhysRev.140.A1133}. This meant that a large portion of the kinetic energy could be calculated precisely, and the remainder was combined with the other unknown parts of the functional, the \emph{exchange and correlation} term. The Kohn-Sham equations are:
\beq
\left(T+V_{KS}\right)\ket{\phi_i}=\varepsilon_i\ket{\phi_i}.
\label{ks-eq}
\eeq
Here $T$ is the electron kinetic energy operator, $V_{KS}$ the Kohn-Sham potential, and $\varepsilon_i$ and $\ket{\phi_i}$ the energy and wave function of the $i$-th Kohm-Sham orbital. A drawback is that the exchange-correlation part of $V_{KS}$ is unknown and needs to be approximated. The wide adoption of DFT that we see today has depended on the development of reliable approximations to the exchange and correlation term~\cite{PhysRevLett.77.3865}. 

\section{STRUCTURES FROM FIRST PRINCIPLES}

The computational discovery of materials with previously unknown structures became practical with the introduction in 2006 of approaches to general first principles structure prediction~\cite{needs2016perspective}. These evolutionary~\cite{oganov2006crystal}, and  random structure searching~\cite{pickard2006high} approaches employed pragmatic strategies for exploring low lying configurations of the DFT energy landscapes generated by state-of-the-art plane wave and pseudopotential codes~\cite{clark2005first,kresse1996efficient}.

The repeated stochastic generation of structures, followed by careful DFT based relaxations to the nearby local minima of the Born-Oppenheimer potential, is the starting point for successful first principles approaches to structure prediction. If no other steps are taken, this is known as \emph{ab initio} random structure searching (AIRSS) and it benefits from parallelism and broad exploratory searches. A particular emphasis is placed on the generation of \emph{sensible} initial structures, where chemical ideas such as coordination, distances, units and symmetry are imposed~\cite{pickard2011ab}. Evolutionary~\cite{oganov2006crystal} and swarm approaches~\cite{wang2012calypso} build subsequent moves  on what has already been learned about the energy landscape, trading some simplicity, parallelism and exploratory power for a greater exploitation of this hard won information. The different approaches appear to be complementary, and the combined application of random search and swarm based searches have been particularly powerful in the study of the hydrides~\cite{li2016dissociation,peng2017hydrogen}. In combination with general purpose plane wave DFT codes~\cite{clark2005first,kresse1996efficient,Giannozzi_2009}, databases of reliable potentials covering the entire periodic table~\cite{lejaeghere2016reproducibility}, and the arrival of commodity multi-core CPUs, first principles structure prediction has now become widespread, and almost routine~\cite{oganov2019NRM}.

The same trends in software and computer architecture have led to high throughput approaches to materials informatics~\cite{jain2016computational}. These, at least initially, depend on the availability of curated databases of crystal structures. However, they have not yet proven to be of use to the study of the dense hydrides -- whose crystal structures are typically not to be found in existing databases. Indeed, even for those structure prototypes which might be available in a database, using modern structure prediction methods it can be easier, faster and more reliable, to \emph{rediscover} the structures, rather than draw candidates from a database, relax and compute their ground states energies from first principles, and sort among them.

\begin{figure}[htbp]
\includegraphics[width=0.66\textwidth]{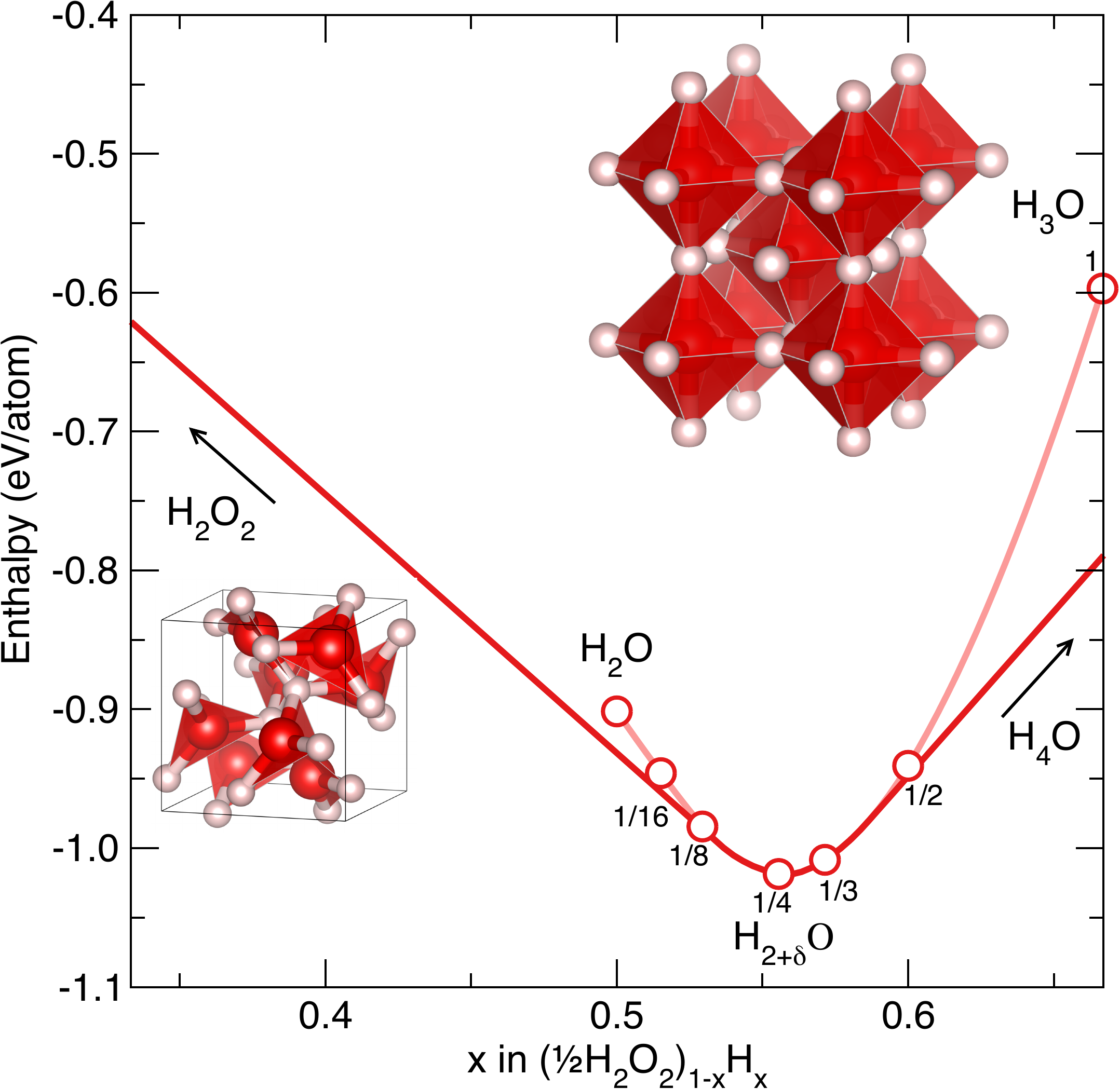}
\caption{Convex hull for the hydrogen-oxygen system at 6 TPa. Under these extreme pressures water (H$_2$O) decomposes into hydrogen peroxide (H$_2$O$_2$) and a hydrogen rich phase (H$_{2+\delta}$O). At $\delta=1$ the H$_3$O structure is the same as the superconducting $Im\bar{3}m$ H$_3$S phase. The stable compositions lie on the convex hull, and H$_3$O is seen to be unstable to a density of states lowering change in composition~\cite{pickard2013decomposition}.}
\label{fig4}
\end{figure}

There have been many striking applications of first principles structure prediction, in particular to high pressure phase transitions~\cite{zhang2017materials}. In the absence of experimentally derived information, structure prediction has provided the most reliable microscopic models of dense hydrogen itself. Using random search, a convincing model of phase III was introduced~\cite{pickard2007structure}, which exhibited the observed strong IR activity. \emph{Mixed} phases were also encountered in the search, and these anticipated the experimental discovery of phase IV~\cite{eremets2011conductive,howie2012mixed}. As an end-member, good models for the high pressure phases of hydrogen have proved important in the search for the binary dense hydrides. Using Maxwell constructions, or convex hull plots (see Fig \ref{fig4}) the stability of these binary (or ternary and above) hydrides can be straightforwardly assessed~\cite{feng2008emergent}. 

\section{SUPERCONDUCTORS FROM FIRST PRINCIPLES}
\label{sec-sc-ff}

In the known superconducting hydrides, the coupling mechanism driving the condensation of the Cooper pairs is the well-known electron-phonon interaction: they are conventional superconductors. This means that it is possible to perform first principles calculations of their superconducting critical temperatures using established theoretical and computational approaches. Exploiting the dramatic increase in available computational power, these first principles calculations have  been central to the characterization and understanding of the properties of superconducting hydrides, and importantly, to predict new high-$T_c$ compounds. 

Once the crystal structure for a given material is known, three basic ingredients are required to calculate its $T_c$ within DFT: the Kohn-Sham  energies $\varepsilon_i$ and wave functions $\ket{\phi_i}$, where $i$ labels a given electronic state; the phonon frequencies $\omega_{\mu}$, with a mode index $\mu$; and the electron-phonon matrix elements~\cite{RevModPhys.89.015003},
\beq
g_{ij}^{\mu}=\bra{\phi_i} \frac{\partial V_{KS}}{\partial u^{\mu}} \ket{\phi_j}.
\label{e-ph-matrix-elem}
\eeq
In the above, $u^{\mu}$ is the atomic displacement according to the normal mode $\mu$. Phonon frequencies are now routinely calculated within the harmonic approximation, truncating the Born-Oppenheimer energy surface at second order. The harmonic force constants are  calculated  by making use of linear response theory~\cite{RevModPhys.73.515} or finite difference approaches~\cite{PhysRevB.75.205413}. The electron-phonon matrix elements are obtained analogously from linear response~\cite{RevModPhys.73.515,PhysRevLett.77.1151} or finite difference methods~\cite{PhysRevB.75.205413,Monserrat_2018}. 

Bringing together the Kohn-Sham energies, phonon frequencies, and electron-phonon matrix elements, the Eliashberg function $\alpha^2F(\omega)$ can be directly evaluated as a phonon density of states weighted by the electron-phonon interaction at the Fermi energy $\varepsilon_F$:
\beq
\alpha^2F(\omega) =  \sum_{ij\mu} |g_{ij}^{\mu}|^2 \delta(\omega-\omega_{\mu}) \delta(\varepsilon_i-\varepsilon_F)
\delta(\varepsilon_j-\varepsilon_F).
\label{eliashberg-eq}
\eeq
This function is central to the prediction of $T_c$ in superconductors. The electron-phonon coupling constant is calculated as
\beq
\lambda = 2\int_0^{\infty} \mathrm{d}\omega \frac{\alpha^2F(\omega)}{\omega},
\label{lambda}
\eeq
and it measures the strength of the attractive interaction between the electrons and the phonons. The semi-empirical McMillan equation~\cite{PhysRevB.12.905}
\beq
k_B T_c = \frac{\hbar \omega_{\mathrm{log}}}{1.2} \exp\left[- \frac{1.04(1+\lambda)}{\lambda-\mu^*(1+0.62\lambda)}\right],
\label{mcmillan}
\eeq
is typically used to predict the critical temperature. The average logarithmic frequency $\omega_{\mathrm{log}}$ can be computed from
\beq
 \omega_{\mathrm{log}}=\exp \left[ \frac{2}{\lambda}\int_0^{\infty} \mathrm{d}\omega \frac{\alpha^2F(\omega)}{\omega} \ln \omega \right],
\label{wlog}
\eeq
and $\mu^*$ is the so-called Coulomb pseudopotential, which accounts for the repulsive electron-electron interaction. The latter is usually taken as a parameter around 0.1, though it can also be explicitly calculated\~cite{doi:10.7566/JPSJ.87.041012}. 

This approach has been successful in accurately computing the $T_c$ of several compounds~\cite{PhysRevLett.87.037001,PhysRevLett.95.237002,PhysRevLett.101.016401}, but it suffers from limitations which are particularly important for the superconducting hydrides. First, the McMillan equation tends to systematically underestimate $T_c$ for strongly coupled superconductors ($\lambda > 1$)~\cite{PhysRevB.12.905}. These difficulties can be overcome by  directly solving the many-body Migdal-Eliashberg equations for the superconducting gap~\cite{ALLEN19831}, or by adopting a density functional theory for superconductors (SCDFT)~\cite{PhysRevLett.60.2430}, which is an extension of DFT accounting for the superconducting state. As an example, the $T_c$ predicted with the McMillan equation for H$_3$S in the cubic $Im\bar{3}m$ phase at 200 GPa is 125 K, whereas the Migdal-Eliashberg equations yield 194 K ($\lambda=1.84$ in this case)~\cite{PhysRevLett.114.157004}. 

\begin{figure}[htbp]
\includegraphics[width=0.85\textwidth]{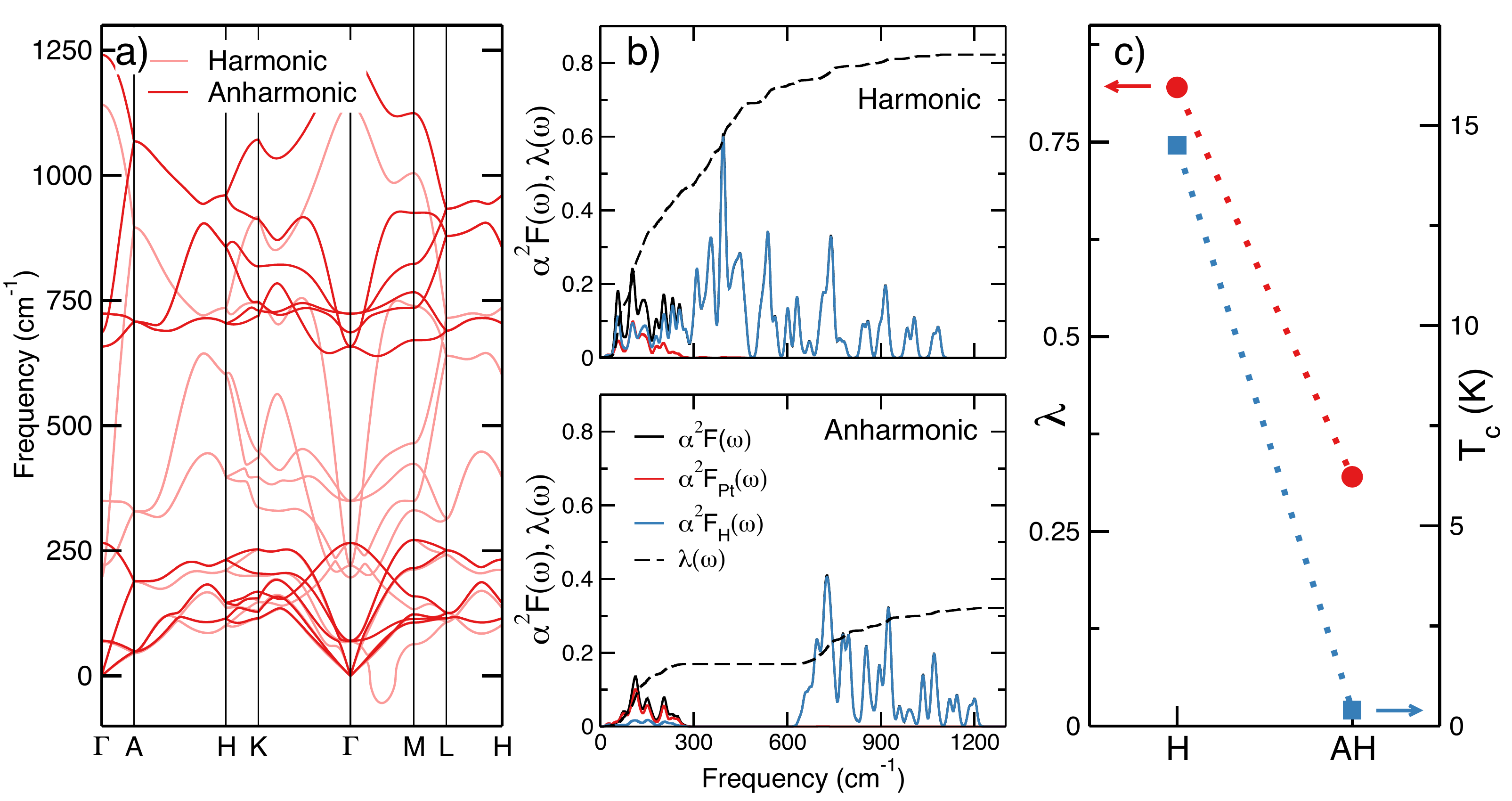}
\caption{(a) Phonon spectra of PtH in the hcp structure at 100 GPa for the harmonic and anharmonic cases. (b) Eliashberg function $\alpha^2F(\omega)$ and the integrated electron-phonon coupling $\lambda(\omega) = 2\int_0^{\omega} \mathrm{d}\Omega \alpha^2F(\Omega)/\Omega$ in the harmonic and anharmonic cases. The decomposition of $\alpha^2F(\omega)$ into H and Pt contributions is included. (c) The calculated $\lambda$ and $T_c$ in the harmonic (H) and anharmonic (AH) calculations~\cite{PhysRevB.89.064302}.}
\label{fig5}
\end{figure}

A second important limitation is the breakdown of the harmonic approximation used to calculate the phonon frequencies. The electron-phonon coupling constant strongly depends on the phonon frequencies: $\lambda\sim\sum_{\mu}1/\omega^2_{\mu}$. If anharmonic effects significantly renormalise the phonon frequencies, $\lambda$ can be substantially modified, and, as a result, so can $T_c$. Because of the low mass of hydrogen and its large quantum fluctuations from equilibrium, substantial anharmonic corrections to $T_c$ have been predicted in many superconducting hydrides and some candidate phases of hydrogen~\cite{PhysRevLett.114.157004,PhysRevB.82.104504,PhysRevLett.111.177002,PhysRevB.89.064302,errea2016quantum,PhysRevB.93.094525,Borinaga_2016}, though not for all~\cite{PhysRevB.93.174308}. In Figure \ref{fig5} we illustrate the effect of anharmonicity with the calculation performed in Ref.~\cite{PhysRevB.89.064302} for PtH at 100 GPa in the hexagonal closed-packed (hcp) structure which has been synthesised experimentally at lower pressures~\cite{PhysRevB.83.214106}. There is strong anharmonic hardening of the phonon energies in this compound, which is mostly associated with the hydrogen related modes, and a consequent suppression of $\lambda$ and $T_c$, by greater than an order of magnitude for the latter.  

\section{THE ROUTE TO SUPERCONDUCTING HYDRIDES}

\begin{figure}[htbp]
\includegraphics[width=0.75\textwidth]{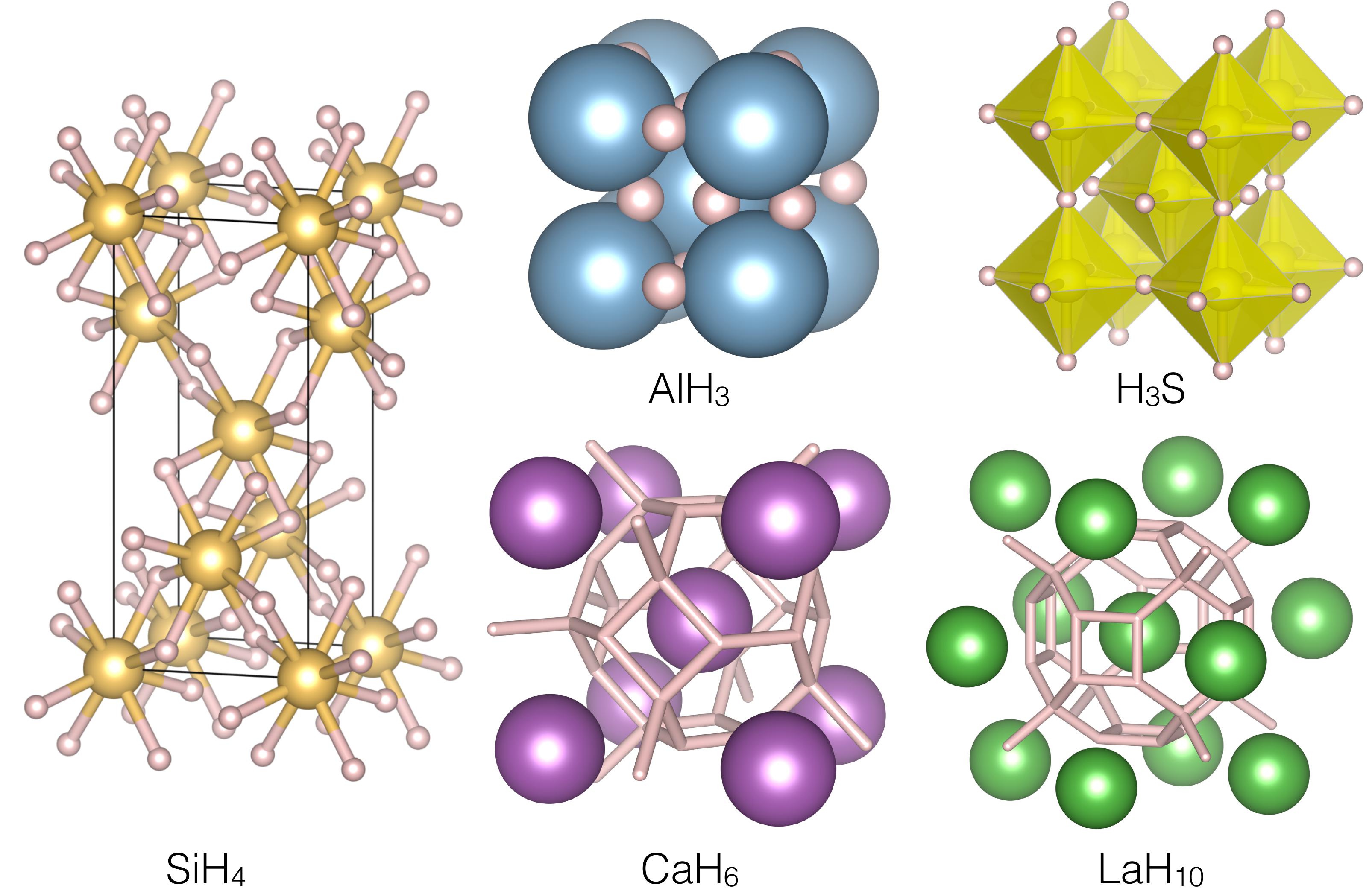}
\caption{Gallery of key hydride structures. SiH$_4$ and AlH$_3$ at 100GPa, and H$_3$S, CaH$_6$ and LaH$_{10}$ at 200GPa. The H$_3$S structure consists of interpenetrating ReO$_3$ lattices. CaH$_6$ and LaH$_{10}$ exhibit a striking hydrogen framework structure, and have been referred to as clathrates, or sodalite-like.}
\label{fig6}
\end{figure}

\begin{figure}[htbp]
\includegraphics[width=0.85\textwidth]{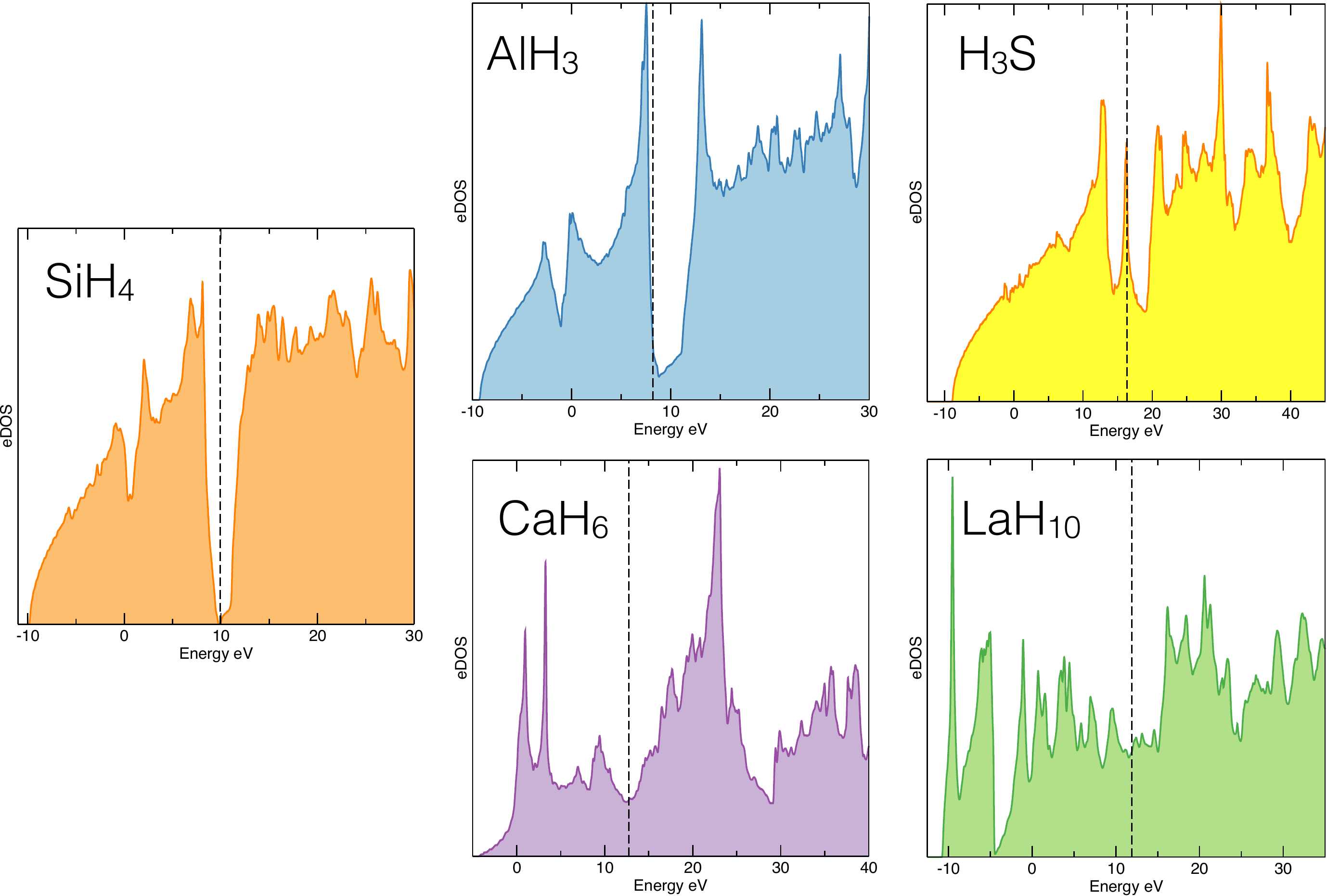}
\caption{The electronic density of states (eDOS) for SiH$_4$~\cite{pickard2006high} at 100GPa, AlH$_3$\cite{pickard2007metallization,goncharenko2008pressure} at 100GPa, H$_3$S~\cite{Duan2014} at 200GPa, CaH$_6$~\cite{wang2012superconductive} at 200GPa and LaH$_{10}$\cite{peng2017hydrogen,liu2017potential} at 200GPa. The Fermi level, $\varepsilon_f$, is indicated by a vertical dashed line. Note that a small gap opens up for SiH$_4$, and so it is not expected to superconduct~\cite{pickard2006high}. AlH$_3$ is metallic~\cite{pickard2007metallization,goncharenko2008pressure}, but anharmonicity destroys high-$T_c$~\cite{PhysRevB.82.104504}. H$_3$S exhibits a  peak in the eDOS close to $E_f$, pointing to its remarkable properties. Both CaH$_6$ and LaH$_{10}$ are good metallic hydrides.}
\label{figX}
\end{figure}

Ashcroft and Hoffmann's first suggestion that compressed silane (SiH$_4$) might take us to metallic superconducting hydrogen~\cite{PhysRevLett.96.017006}, at lower pressures than pure H$_2$, was backed up by first principles computations of its expected properties~\cite{PhysRevLett.96.017006}. The structures investigated were derived largely from chemical intuition, and using the newly developed first principles structure prediction techniques it was quickly shown that there were more stable phases which were expected to be semiconducting, and hence poor candidates for superconductivity~\cite{pickard2006high}. Experiments confirmed these structural predictions, and the $T_c$ was found to be low~\cite{eremets2008superconductivity}. Methane (CH$_4$), and germane (GeH$_4$) were suggested~\cite{gao2008superconducting} but they also did not exhibit high temperature superconductivity. The hydrogen storage materials (LiBH$_4$, NaBH$_4$, NH$_3$BH$_3$, Si(CH$_3$)$_4$) were obvious candidates, given their high hydrogen content, but they resisted metallisation to high pressures. One of them, aluminum hydride (AlH$_3$) was found to metallise, both theoretically~\cite{pickard2007metallization} and experimentally~\cite{goncharenko2008pressure}, but did not superconduct at the 20K or so that it was computed to do so. This was later explained to be a result of strong anharmonic effects~\cite{PhysRevB.82.104504}. Despite these disappointments, some groups persisted and went on to make remarkable predictions, most notably that CaH$_6$ would have a $T_c$ of 235 K at 150 GPa~\cite{wang2012superconductive}. The structures of some of these compounds are shown in Figure \ref{fig6} and their electronic density of states (eDOS) in Figure \ref{figX}.

\section{DISCOVERY OF SUPERCONDUCTIVITY IN HYDROGEN SULFIDE}

In the face of the early failures, the experimental quest continued. H$_2$S was selected as it is widely available, and had been predicted to superconduct with a $T_c$ around 80 K at high pressure~\cite{li2014metallization} by a group with a good track record, that successfully anticipated the emergence of transparent sodium under compression~\cite{ma2009transparent}. It was a good choice, as superconductivity with $T_c$ 50-60 K was found, in reasonable agreement with theory. Already a record for conventional superconductors, further inspection revealed  a strong increase in $T_c$ with pressure, up to about 150 K. Serendipitously, it was noticed that not only pressure but also increasing temperature led to $T_c$ to soar. The sample was then deliberately heated, and $T_c$ further increased, stabilizing at around 200 K. It was suspected that H$_2$S disproportionated with temperature, likely transforming to H$_3$S plus sulphur. A similar decomposition had been predicted in water (H$_2$O) at terapascal pressures (see Figure \ref{fig4}), a chemical analogue for H$_2$S.

This observed superconductivity was characterized by zero resistance, a shift of $T_c$ to lower temperatures with applied magnetic field, and a strong isotope effect (through the replacement of H$_2$S with D$_2$S) that pointed to conventional superconductivity. Crucially, the Meissner effect was observed. This required the development of a new high pressure technique: the use of a sensitive SQUID (superconducting quantum interference device) magnetometer. In order to accomodate a SQUID, DACs smaller than 9 mm in diameter were required. Such tiny DACs had previously worked up to 15 GPa, and fortunately they also did so at 200 GPa, providing the final convincing evidence of superconductivity~\cite{Drozdov2015}. The fact that independent calculations suggested, almost at the same time that (H$_2$S)$_2$H$_2$ may be a high-$T_c$ compound~\cite{Duan2014} provided further support. The predictions and experiments were  consistent. 

\section{A THEORETICAL UNDERSTANDING EMERGES}

This experimental discovery of superconductivity at 203 K in hydrogen sulfide~\cite{Drozdov2015} stimulated further intense theoretical work, which has proven to be crucial to the full characterization and understanding of its properties. Variable stoichiometry crystal structure predictions clearly determined that H$_2$S is not thermodynamically stable above 50 GPa and that it decomposes mainly into H$_3$S and S, as suggested by the experiments, although other decomposition mechanisms have also been considered~\cite{PhysRevLett.114.157004,li2016dissociation,PhysRevB.91.180502}. Among all the possible compounds resulting from the decomposition, first principles calculations soon determined that only H$_3$S could provide such an extraordinary $T_c$~\cite{PhysRevLett.114.157004,li2016dissociation,PhysRevB.93.094525,PhysRevB.91.224513,Flores-Livas2016,errea2016quantum}. All other possibilities yielded values of $\lambda$ that were too low. This picture that H$_2$S decomposes yielding H$_3$S is further supported by the fact that the rise in $T_c$ with increasing pressure observed~\cite{Drozdov2015} is consistent with the theoretical $T_c$ provided by a gradual transformation of H$_2$S into H$_3$S~\cite{PhysRevLett.117.075503}.

\begin{figure}[htbp]
\includegraphics[width=0.75\textwidth]{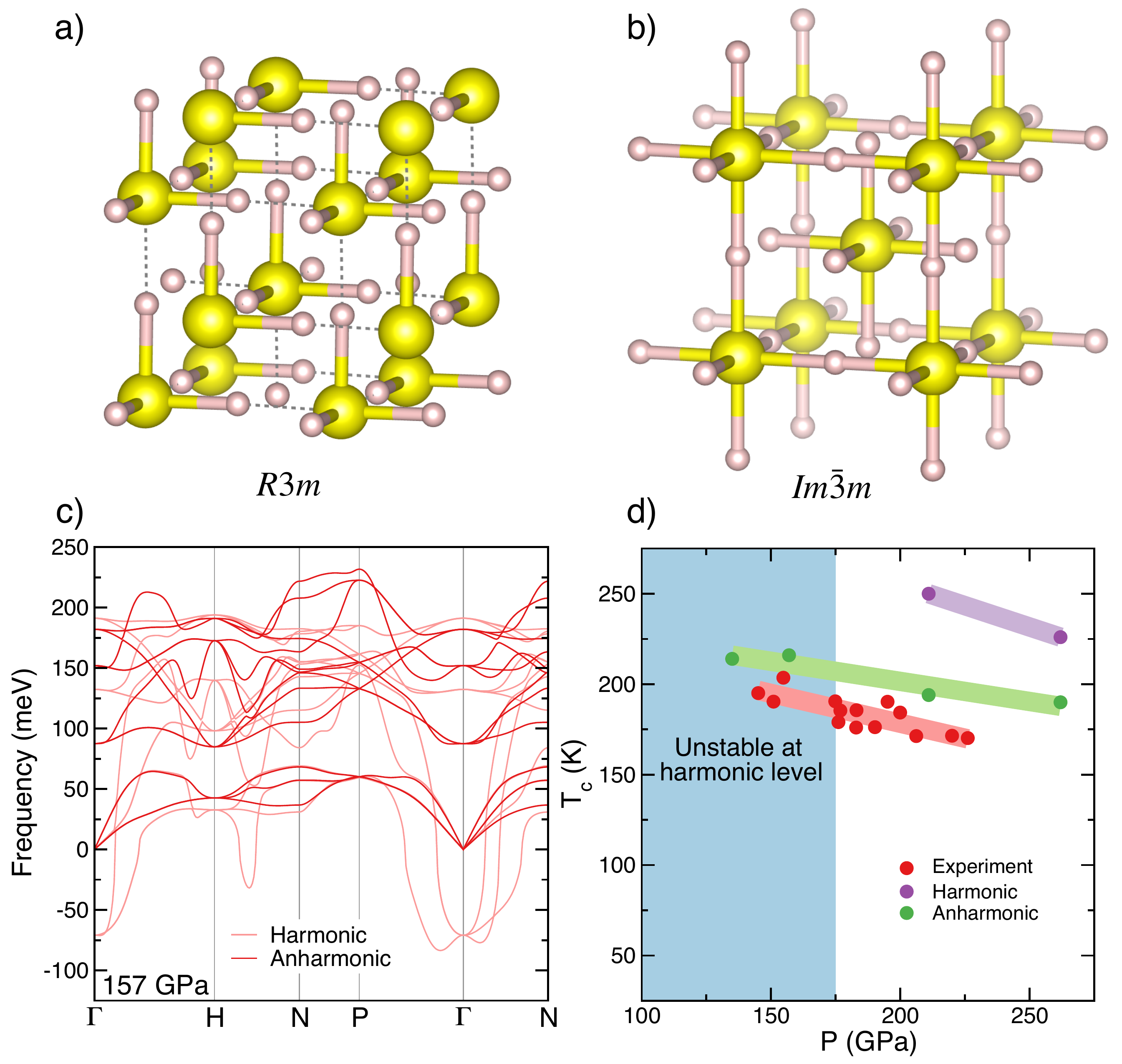}
\caption{Crystal structure of (a) $R3m$ and (b) $Im\bar{3}m$ phases of H$_3$S. (c) Phonon spectra of $Im\bar3m$ H$_3$S at 157 GPa and (d) $T_c$ as a function of pressure of $Im\bar{3}m$ H$_3$S calculated in the harmonic approximation and including anharmonicity. In (d) the blue region describes the pressures at which $Im\bar{3}m$ H$_3$S is not stable in the harmonic approximation. The experimental data for H$_3$S in Ref.~\cite{Drozdov2015} is provided for reference. The theoretical data for this figure is from Ref.~\cite{errea2016quantum}.}
\label{fig8}
\end{figure}

The phase sequence predicted for H$_3$S  suggests a $Cccm$~\cite{Duan2014} or $C2/c$~\cite{li2016dissociation} structure below 112 GPa, both formed of H$_2$S and H$_2$ units that cannot explain the large $T_c$, a rhombohedral $R3m$ phase between 112 GPa and approximately 175 GPa, and a cubic $Im\bar{3}m$ phase above~\cite{Duan2014,li2016dissociation}. As shown in Figure \ref{fig8}, the H atoms in the $Im\bar{3}m$ phase sit exactly halfway between two sulfur atoms forming a structure with full cubic symmetry. At lower pressures, the hydrogen atoms move to an off-centre position, forming a short H$-$S covalent bond and a longer H$\cdots$S hydrogen bond, lowering the symmetry to $R3m$. The displacive transition from  $Im\bar{3}m$ to  $R3m$ is driven by the softening of a phonon mode at the $\Gamma$ point. 

The above sequence of phases was determined neglecting the contribution of the ionic fluctuations to the energy, the quantum zero point energy, and so a classical prediction. As discussed in Sec. \ref{sec-sc-ff}, quantum fluctuations mean that hydrogen atoms vibrate with a large amplitude from equilibrium even at absolute zero, which can lead to a substantial anharmonic renormalization of the phonon frequencies. As shown in Ref.~\cite{errea2016quantum}, once the zero point energy is included in the calculations, the $R3m$ is no longer the ground state structure below 175 GPa, the cubic $Im\bar{3}m$ is favorable even if it is dynamically unstable in the harmonic approximation. Anharmonicity stabilizes the phonons of the cubic phase and yields $T_c$ values in agreement with experiments (see Figure \ref{fig8})~\cite{errea2016quantum}. The highest temperature at which superconductivity is observed in H$_3$S occurs in a structure with hydrogen bonds symmetrized by quantum effects. Once these quantum anharmonic effects are correctly included, the transition between the $Im\bar{3}m$ and  $R3m$ phases is estimated to be between 91 and 114 GPa~\cite{PhysRevB.97.214101}.

\section{FURTHER EXPERIMENTAL CHARACTERIZATION}

Early X-ray diffraction (XRD) measurements~\cite{Einaga2016}, performed on a compressed H$_2$S sample, confirmed its decomposition yielding H$_3$S with a bcc arrangement of the S atoms, but could not distinguish between the $R3m$ and $Im\bar{3}m$ phases. Hydrogen atoms are very weak scatterers and their position cannot easily be determined by XRD. $T_c$ was measured on further pressure release and showed a pronounced kink at 150 GPa, which could signal the occurrence of the $Im\bar{3}m \to R3m$ transition~\cite{Einaga2016}. Two more recent experimental studies synthesized clean H$_3$S by annealing a sulphur sample in a DAC loaded with H$_2$ gas. Goncharov et al.~\cite{PhysRevB.95.140101} confirmed the theoretically predicted $Cccm \to R3m \to Im\bar{3}m$ structural sequence. The cubic $Im\bar{3}m$ was directly synthesized at high pressure, and subsequent pressure release led to the appearance of a rhombohedral distortion compatible with the $R3m$ phase at 140 GPa, which remained metastable down to 70 GPa, where it transformed upon annealing to the $Cccm$ structure. The observed rhombohedral distortion is much larger than that expected theoretically~\cite{PhysRevB.97.214101}, which could be due to  slight non-hydrostatic conditions in the DAC. This raises hopes of preserving the cubic $Im\bar{3}m$ structure and its large $T_c$ to  even lower pressures. By annealing H$_2$ and S at lower pressures instead, Guigue et al.~\cite{PhysRevB.95.020104} were only able to synthesize $Cccm$ H$_3$S, which remained metastable up to 160 GPa. Taken together, these results suggest that the transition between the $Cccm$ phase to the $R3m$ or $Im\bar{3}m$ phases is strongly affected by large kinetic barriers. This observation has important implications for the predictability of high pressure hydrides.

The superconducting state of H$_3$S has been further characterized by optical and magnetic measurements~\cite{Capitani2017,Troyan1303,2019arXiv190111208M}. Capitani et al.~\cite{Capitani2017} found evidence for the presence of a large electron-phonon mediated  superconducting gap in reflectivity measurements, which was in agreement with the reflectivity calculated with the anharmonic $\alpha^2F(\omega)$ in Ref.~\cite{errea2016quantum}. Recent magnetic measurements up to 65 T at 155 GPa show a critical magnetic field consistent with a strongly coupled superconductor with $\lambda\sim 2$, a value in agreement with first principles computations~\cite{errea2016quantum}.   

\begin{figure}[htbp]
\includegraphics[width=0.75\textwidth]{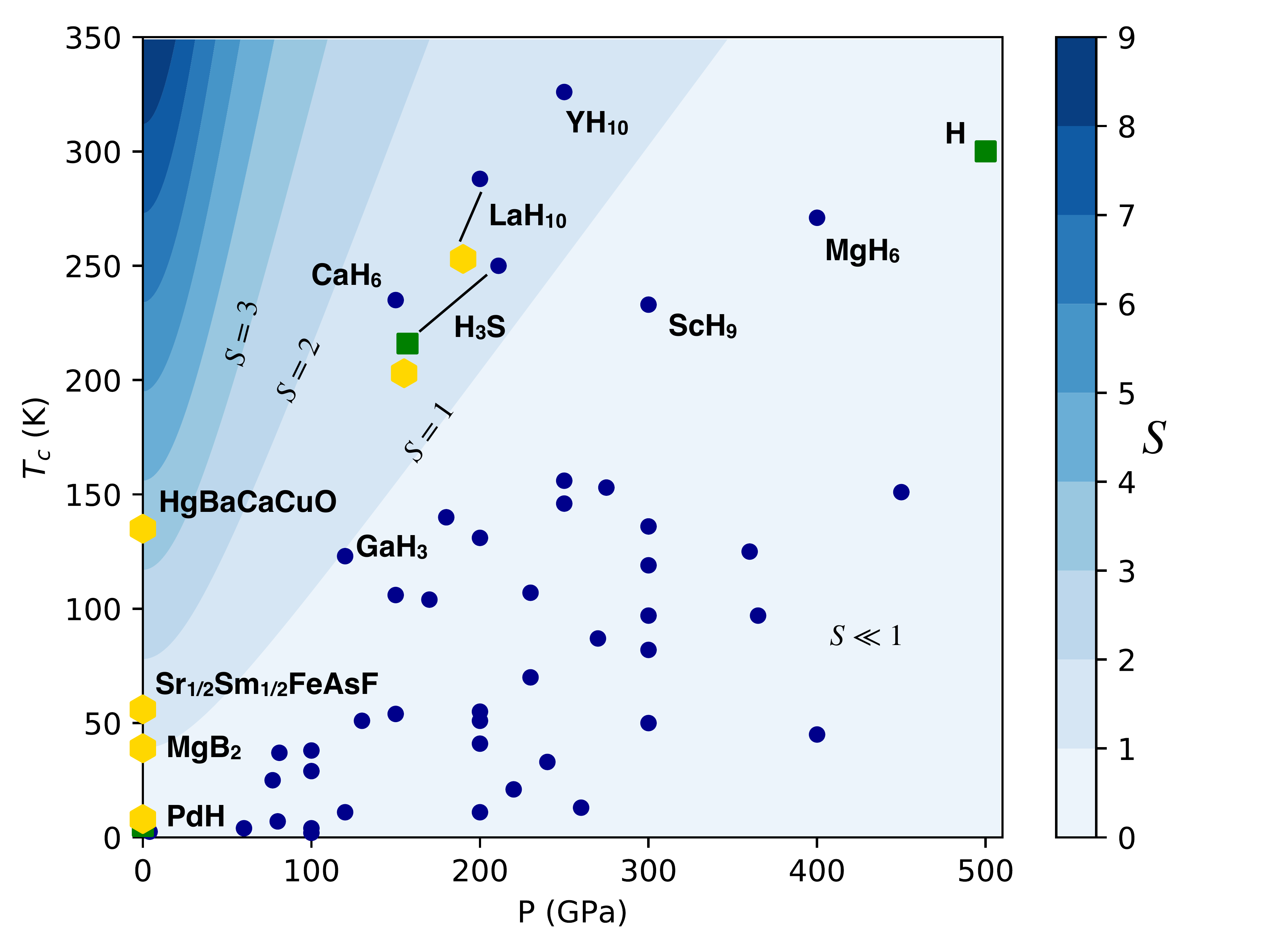}
\caption{Superconducting critical temperature, $T_c$, as a function of the pressure at which it has been calculated or measured for different hydrides. Yellow hexagons correspond to experimental measurements~\cite{PhysRevB.10.3818,Drozdov2015,PhysRevLett.122.027001,Schilling1993,Wu_2009,Nagamatsu2001}. Green squares correspond to first principles calculations including anharmonic effects~\cite{PhysRevLett.114.157004,PhysRevLett.111.177002,PhysRevB.93.174308}. All other small circles correspond to predictions at the harmonic level as summarized in  Table 1 of Ref.~\cite{2018arXiv180600163B}, except for H$_3$S~\cite{errea2016quantum}. The coloured contours correspond to the figure of merit $S$ proposed in Eq. \ref{fig_merit_s}.}
\label{fig9}
\end{figure}

\section{LANTHANUM HYDRIDES AND BEYOND}

The discovery of superconductivity above 200 K in H$_3$S in 2015 showed that hydrides can indeed be high-$T_c$ superconductors. At the same time it clearly illustrated how fruitful the combination of theory and experiment can be in the characterization and understanding of these materials. Soon, evidence that PH$_3$ superconducts at 100 K and 200 GPa was reported~\cite{2015arXiv150806224D}. First principles calculations, however, show that PH$_n$ compounds are not thermodynamically stable with respect to the decomposition into phosphorus and hydrogen, suggesting that superconductivity might have occurred in a metastable state in a compound with an unknown stoichiometry~\cite{PhysRevB.93.020508}.  

More recently, evidence for  superconducting transitions as high as above 250 K have been reported in a lanthanum hydride at around 150-200 GPa by two independent groups~\cite{PhysRevLett.122.027001,2018arXiv181201561D,2018arXiv180807039D}. 
The synthesis was achieved by directly annealing in the DAC La and H$_2$ gas~\cite{2018arXiv181201561D,2018arXiv180807039D} or using BH$_3$NH$_3$ as the hydrogen source. The latter option dramatically simplifies the experiment as only solid samples are used, even if the synthesis is less well controlled. In any case, a severe experimental difficulty encountered is that phases with different structures and stoichiometry are synthesized at nearly the same pressure-temperature conditions, and the final product depends on the kinetics of the transformations. Based on the volume per formula unit, a stoichiometry of around LaH$_{10}$ has been estimated~\cite{2018arXiv181201561D,doi:10.1002/anie.201709970}. The most probable candidate for such an extraordinary value of $T_c$ is a hydrogen clathrate structure with LaH$_{10}$ stoichiometry and space group $Fm\bar{3}m$ (see Figure \ref{fig6}), previously predicted to be a high-$T_c$ superconductor by first principles calculations~\cite{peng2017hydrogen,liu2017potential}. For this structure a pure superconducting metallic hydrogen lattice exists, to which the host La atoms donate electrons~\cite{liu2017potential}. XRD measurements are consistent with this phase~\cite{2018arXiv181201561D,doi:10.1002/anie.201709970}. Nevertheless, different values of $T_c$ have been observed~\cite{2018arXiv181201561D,2018arXiv180807039D} and XRD experiments find very different phases for the hydrogen and deuterium compounds~\cite{2018arXiv181201561D}. Further theoretical calculations that accurately account for quantum anharmonic effects are thus needed to clarify the phase diagram and the superconducting nature of these hydrides.

By now there are a very large number of hydrides that have been predicted theoretically to be thermodynamically stable and exhibit a high $T_c $. Figure \ref{fig9} summarizes many of these predictions, which are discussed in detail in Ref.~\cite{2018arXiv180600163B}. Even if such predictions might once have been unbelievable, ignored or criticized by part of the superconducting community~\cite{Hirsch_2009}, it is now clear that there is plenty of room for further groundbreaking experimental discoveries, although experimental progress is slow as compared to theory and computation. We could  ask ourselves, why should this be the case? One obvious reason is the difficulty of the experiments. It could be because the synthesis of these dense hydrides is hindered by large kinetic barriers due to the making and breaking of hydrogen dimers. Or simply because many of these predictions are inaccurate, in particular, because the quantum nature of the hydrogen atoms is usually neglected, or because of the intrinsic limitations of DFT, or the extensiveness of the structural searches.

\section{DISCUSSION}

There is continuing interest in the experimental results for the lanthanum hydrides, and we can expect further experimental investigations. At the same time theoretical groups will attempt to refine our microscopic understanding of this system, in particular exploring the impact of the quantum dynamical behaviour of the protons on structure, stability and superconductivity. And no doubt the computational search for new candidates will continue. It is possible that the true structures of the hydrides are more complex than the fairly small unit cells typically investigated, and our understanding of the binary hydrides may be refined. Beyond that, the ternaries beckon.

Many questions, as well as challenges and opportunities remain. What do these recent successes in the superconducting hydrides mean for the dream of the discovery of high temperature superconducting materials? Is there a limit to how high $T_c$ can be in these conventional superconductors, and how does that limit depend on pressure, composition and structure? Intuitively there should be a limit, after which superconductivity is out competed by structural distortion, compositional change, or other electronic or magnetic phases. Efforts in quantifying this will be valuable. Is extremely high pressure essential, or might these results be opening our eyes to the possibility of room temperature superconductivity under ambient, or close to ambient, conditions? 

The wide range of $T_c$ values predicted in superconducting hydrides~\cite{2018arXiv180600163B}, from few kelvins to above 300 K (see Figure \ref{fig9}), suggests that Ashcroft's remarkable idea~\cite{ashcroft2004hydrogen,ashcroft1968metallic} was too general and that high $T_c$ in hydrides is not just related to the Debye temperature being large. A strong electron-phonon coupling is also required. The range of $\lambda$ in the hydrides is consequently also very large, with values from around 0.4 in PdH~\cite{PhysRevLett.111.177002} to $\lambda$ about 2 in H$_3$S~\cite{PhysRevLett.114.157004,errea2016quantum}. A clear understanding of when a hydride yields a large $\lambda$ will turn out crucial to clarify the prospects of superconducting hydrides.  

Superconductivity in the hydrides is forcing is to ponder what we mean by \emph{room temperature}. One  definition might be 0$^oC$ (273K), but we can all agree that this would be a very cold room. Maybe 290K is more reasonable, but we should remember that for technological applications the superconducting material would need to operate at well below $T_c$.

Of course, no room can be held at the megabar (100 GPa) pressures currently required to force the hydrides into the superconducting state. Indeed, as we have seen, these very high pressures mean that only very few experimental groups can participate in the search for new superconducting hydrides. It is essential that the pressures required are reduced. This could be promoted by computational predictions which seek a compromise and balance the pressure required with the $T_c$ predicted, rather than simply trying to maximise $T_c$ with no regard to the experimental conditions required. To this end we propose a figure of merit, $S$, which make the compromise explicit:
\begin{equation}
S=\frac{T_c}{\sqrt{T_{c,\rm MgB_2}^2+P^2}},
\label{fig_merit_s}
\end{equation}
where the temperatures are in Kelvin, and the pressures in GPa. This sets MgB$_2$ (a high temperature conventional superconductor at ambient pressures, with technological applications) to have $S$(MgB$_2$)=1. On this scale a putative superconductor with a $T_c$ of 390K at ambient pressures (and so could be used without cooling or compression in a wide range of terrestrial conditions) would score a perfect $S$($\star$)=10. The current megabar superconducting hydrides have lower values (see Figure \ref{fig9}), reflecting the very high pressures required to achieve the superconducting phases -- with both $S$(H$_3$S) and $S$(LaH$_{10}$)=1.3. A superconductor with a $T_c$ of around 1000K at 100 GPa would score nearly 10 on this scale, which captures the astonishment that such a result would generate, and $S$(HgBaCaCuO)= 3.5 reflecting the Nobel Prize worthy discovery of the cuprates.

An enduring puzzle is the disparity between the number of the theoretically predicted superconductors that now populate the literature, and the few that have been experimentally realized. As mentioned above, this may partly be due to the relatively few groups that can currently perform the necessary experiments. But it is not the whole story, and it would be helpful if theoretical predictions could comment on the likely ease (or otherwise) of experimental synthesis. As the experimental evidence reviewed here suggests, large kinetic barriers appear to be hindering the synthesis of the superconducting hydrides.  

As was seen with the early investigations of compressed silane and the decomposition of H$_2$O, more stable structures or compositions will typically have lower densities of electronic states at the Fermi level, reducing the prospects of high temperature superconductivity. This leads to a potential bias in the predictions towards higher $T_c$. It is not possible to guarantee that a ground state structure has been identified in any stochastic search, but searches halted too soon (for example when a pleasing result has been obtained) are potentially unreliable.

On the computational side, the high throughput sweep of databases, or stochastic searches have become relatively routine. However, the computation of $T_c$ has not. The most reliable results take care~\cite{zarifi2018crystal}, and very large computational resources. This becomes even more the case if  anharmonic effects, that we have seen are important for hydrogen containing compounds, are to be computed. Progress in this direction, in particular the automation of the computations, would advance the field considerably.

Unconventional superconductivity may also be encountered in high pressure experiments. Unfortunately, predictive computational methods are not currently helpful in this case. Should quantitative theories emerge for  unconventional superconducting states, we could look forward to the same fruitful symbiosis between theory, computation and experiment which has been so successful for the superconducting hydrides.

\section{CONCLUSION}

The existence of high temperatures superconductivity in metallic hydrogen, or hydrogen rich compounds, has been long been theoretically discussed and in 2014 superconductivity was discovered in compressed hydrogen sulfide at 203K and around 150 GPa. In 2018 superconductivity was observed in compressed lanthanum hydride at above 250K and around 200 GPa by two independent groups. First principles structure and superconductivity predictions have played a crucial role in guiding these experimental discoveries. A detailed theoretical picture of the superconducting mechanism has emerged for H$_3$S, and it is expected to do so for LaH$_{10}$

Experiments involving hydrogen at megabar pressures are extremely challenging. In view of the recent discoveries, the theoretical effort will continue in the coming years with the hope of leading the design of more accessible high temperature superconducting hydrides. However, mindful of the apparently singular success of MgB$_2$, are we at risk of being misled that there are many more such superconductors to be discovered at ambient pressure? Can a similar combination of theory and computation lead the discovery of new high $T_c$ superconductors beyond the hydrides at technologically relevant conditions?      

\section*{ACKNOWLEDGMENTS}
CJP is supported by the Royal Society through a Royal Society Wolfson Research Merit award. IE has received funding from the European Research Council (ERC) under the European Union’s Horizon 2020 research and innovation programme (grant agreement No 802533), and the Spanish Ministry of Economy and Competitiveness (FIS2016-76617-P). MIE thanks the Max Planck community for invaluable support, and U. P\"oschl for the constant encouragement. We thank Joseph Nelson for a careful reading on the manuscript.

\bibliography{hydrides}

\end{document}